\documentclass{aa}
\bibpunct{(}{)}{;}{a}{}{,} 
\usepackage{natbib}
\usepackage{amsmath}
\usepackage{physics}
\usepackage{graphicx}
\usepackage{txfonts}

\begin{document}

    \title{Effects of an eccentric inner Jupiter on the dynamical evolution \\
           of icy body reservoirs in a planetary scattering scenario
   }

   \author{M. Zanardi\inst{1,2}\thanks{mzanardi@fcaglp.unlp.edu.ar},
           G. C. de El\'\i a\inst{1,2},
           R. P. Di Sisto\inst{1,2},
           S. Naoz\inst{3},
           G. Li\inst{4},
           O. M. Guilera\inst{1,2},
           \and
           A. Brunini\inst{5}
          }

   \offprints{M. Zanardi
    }

   \institute{Instituto de Astrof\'{\i}sica de La Plata, CCT La Plata-CONICET-UNLP \\
   Paseo del Bosque S/N (1900), La Plata, Argentina
   \and Facultad de Ciencias Astron\'omicas y Geof\'\i sicas, Universidad Nacional de La Plata \\
   Paseo del Bosque S/N (1900), La Plata, Argentina
   \and Department of Physics and Astronomy, University of California, Los Angeles, CA 90095, USA
   \and Harvard-Smithsonian Center for Astrophysics, The Institute for Theory and Computation, \\
   60 Garden Street, Cambridge, MA 02138, USA
   \and Universidad Nacional de La Patagonia Austral. Unidad Acad\'emica Caleta Olivia \\
   Ruta 3 Acceso Norte, Caleta Olivia (9311), Santa Cruz. CONICET
                }

   \date{Received / Accepted}

\abstract
{}
{We analyze the dynamics of small body reservoirs under the effects of an eccentric inner giant planet resulting from a planetary scattering event around a 0.5 M$_{\odot}$ star.
}
{First, we used a semi-analytical model to define the properties of the protoplanetary disk that lead to the formation of three Jupiter-mass planets. Then, we carried out N-body simulations assuming that the planets are close to their stability limit together with an outer planetesimal disk. In particular, the present work focused on the analysis of N-body simulations in which a single Jupiter-mass planet survives after the dynamical instability event.
}
{Our simulations produce outer small body reservoirs with particles on prograde and retrograde orbits, and other ones whose orbital plane flips from prograde to retrograde and back again along their evolution (``Type-F particles''). We find strong correlations between the inclination $i$ and the ascending node longitude $\Omega$ of Type-F particles. First, $\Omega$ librates around 90$\degr$ or/and 270$\degr$. This property represents a necessary and sufficient condition for the flipping of an orbit. Moreover, the libration periods of $i$ and $\Omega$ are equal and they are out to phase by a quarter period. We also remark that the larger the libration amplitude of $i$, the larger the libration amplitude of $\Omega$. We analyze the orbital parameters of Type-F particles immediately after the instability event (post IE orbital parameters), when a single Jupiter-mass planet survives in the system. Our results suggest that the orbit of a particle can flip for any value of its post IE eccentricity, although we find only two Type-F particles with post IE inclinations $i \lesssim$ 17$\degr$. Finally, our study indicates that the minimum value of the inclination of the Type-F particles in a given system decreases with an increase in the eccentricity of the giant planet.
}
{} 


\authorrunning{M. Zanardi, et al.
               }
\titlerunning{Effects of an eccentric inner Jupiter on the dynamics of icy reservoirs}

\maketitle

\section{Introduction}

In recent years, an enormous diversity of planetary systems has been discovered around stars of different spectral types. In particular, low-mass stars are of relevant interest because they are the most common stars in our stellar neighborhood. In fact, studies developed by \citet{Henry2006} indicate that 72\% of all the stars within 10 pc are M dwarfs. Moreover, as predicted by \citet{Salpeter1955} and \citet{Chabrier2003}, studies of the solar neighborhood have inferred that M dwarfs are 12 times more abundant than G dwarfs.

The number of extrasolar planets discovered to date around stars of different spectral types amounts to 3610 (http://exoplanets.eu). In particular, 3\%-10\%\footnote{This range is due to the incompleteness of the exoplanet catalog.} of the known exoplanets is orbiting around of M-type stars and some of them are gaseous giants with masses greater than 1 M$_{\text {Jup}}$ and with a wide range of eccentricities. In particular, the system Gliese 317 is formed by an M-type star of 0.42 $\pm$ 0.05 M$_{\odot}$ and two giant planets referred as Gliese 317b and Gliese 317c \citep{AngladaEscude2012}. While Gliese 317b has a mass of 1.81 $\pm$ 0.05 M$_{\text{Jup}}$, a semimajor axis of 1.148 au, and an eccentricity of 0.11 $\pm$ 0.05, Gliese 317c could have orbital parameters very different respect to those associated with the gas giants of the solar system. In fact, a possible solution derived by \citet{AngladaEscude2012} suggests that Gliese 317c could be a giant planet of $\sim$ 2 M$_{\text{Jup}}$, with a semimajor axis between 10 au and 40 au, and an eccentricity of 0.81 $\pm$ 0.2.

One of the channels suggested in the literature to explain the highly eccentric orbits of the observed exoplanets is known as planet-planet scattering \citep{Rasio1996,Weidenschilling1996}. This mechanism arises from strong gravitational perturbations among two or more giant planets of a system, leading to strong dynamical instabilities. These instability events produce crossing orbits and successive close encounters between planets, which could derive in collisions between planets, ejection of one or more of them, as well as collisions with the central star. Several works have been developed to explore the stability of multi-planet systems. On the one hand, \citet{Chambers1996}, \citet{Marzari2002}, and \citet{Marzari2014a} analyzed the dynamics of systems of more than two planets around a solar-type star considering gravitational interactions with each other. On the other hand, using hydrodynamic simulations, \citet{Moeckel2008}, \citet{Marzari2010}, \citet{Moeckel2012}, and \citet{Lega2013} studied the evolution of multi-planet systems, simulating dynamical interactions between the planets and with the gas disk.  

Scattering planetary events seem to play a dominant role in systems with giant planets. An interesting result derived by \citet{Chatterjee2008} and \citet{Juric2008} suggests that planet-planet scattering can reproduce the eccentricity distribution of known extrasolar planets for $e \gtrsim$ 0.2 \citep{Butler2006,Udry2007}. According to these works, about 75\% of the known extrasolar systems have undergone a period of large-scale dynamical instability involving planet-planet scattering events. 

Planetary scattering events lead to a violent scenario of evolution which alters not only the planetary architecture but also the reservoirs of small bodies. Thus, it is very interesting to study the dynamical properties of the minor planets populations resulting from the planet-planet scattering mechanism.

From N-body simulations, \citet{Raymond2013} analyzed the formation and survival of compact isotropic planetesimal clouds from planet-planet scattering in systems around solar-type stars. These authors have shown that reservoirs analogous to ``mini Oort clouds'' with spatial scales of 100-1000 au seem to be a natural byproduct of planetary scattering events in systems with giant planets. In particular, \citet{Raymond2013} found that compact isotropic clouds of planetesimals are preferentially produced in systems with nearly equal-mass Jupiter-like planets. An important result derived in that study suggests that those reservoirs analogous to ``mini-Oort clouds'' may be detectable by their dust signatures at far-IR wavelengths with {\it Spitzer}.

At the same time, \citet{Marzari2014b} made use of N-body simulations and explored the impact of planet-planet scattering on the formation and survival of debris disks around Sun-like stars. In that study, the author suggested that two conditions are strongly unfavorable to the survival of these reservoirs: firstly, an extended period of chaotic behavior before the ejection of one planet and, secondly, the insertion of the outer surviving planet in a highly eccentric orbit. In general terms, \citet{Marzari2014b} inferred that systems which underwent an episode of chaotic evolution might have a lower probability of harboring a debris disk.

On the other hand, \citet{Lithwick2011} studied the dynamical evolution of a test particle that orbits a star in the presence of an exterior massive planet, considering octupole-order secular interactions. When the planet's orbit is eccentric, the Kozai mechanism can lead to a dramatic behavior on the test particle's orbit, which can flip from prograde to retrograde and back again, reaching arbitrarily high eccentricities given enough time. At the same time, \citet{Katz2011} developed a similar work concerning extreme eccentricities and inclinations excited by a distant eccentric perturber. In addition, \citet{Li2014} carried out a chaotic analysis of the test particles under influence of a outer eccentric perturber. A detailed discussion about the effects produced by an eccentric outer perturber on a test particle can be found in \citet{Naoz2016}.

The main goal of the present research is to study the final planetary configuration and the dynamical properties of the reservoirs of icy bodies produced from planetary scattering events around stars of 0.5 M$_{\odot}$, in systems starting with three Jupiter-mass planets close to their stability limit. In particular, this work focuses on systems ending up with a single Jupiter-mass planet after the dynamical instability event. The present paper is therefore structured as follows. In Section 2, we present the properties of the protoplanetary disk used for our study, as well as the numerical code adopted for
developing the N-body simulations. A detailed analysis of the results concerning the dynamical features of the planets and small body reservoirs resulting from our numerical simulations are shown in Section 3. Finally, Section 4 presents the discussions and conclusions of our work.

\section{Numerical methods}

To carry out our research, we made use of two different numerical codes. First, we used a semi-analytical code developed by \citet{Guilera2010} to define the properties of the protoplanetary disk able to form three gaseous giants close to their limit of stability. Once characterized the protoplanetary disk, this was adopted to obtain initial conditions to be used in the N-body code known as Mercury \citep{Chambers1999}. This kind of numerical codes results to be very useful in order to analyze the main dynamical mechanisms involved in the evolution of the planets and small body populations of our simulations. Here, we describe the properties of the protoplanetary disk and present the semi-analytical model and the N-body code used to carry out our research. It is worth noting that none of these codes considers relativistic effects.

\subsection{Properties of the protoplanetary disk}
\label{sec:sec2-1}

The initial surface densities of gas $\Sigma_{\text{g}}$ and planetesimals $\Sigma_{\text{p}}$ determine the initial distribution of material along the protoplanetary disk. For simplicity, in this work we considered that these surface densities are given by a smooth power-law,
\begin{eqnarray}
\Sigma_{\text{g}}(R) &=& \Sigma_{\text{g}}^{0} \left(\frac{R}{1~\text{au}}\right)^{-1}~\text{g}~\text{cm}^{-2}, \nonumber \\
\Sigma_{\text{p}}(R) &=& \Sigma_{\text{p}}^{0} \eta \left(\frac{R}{1~\text{au}}\right)^{-1}~\text{g}~\text{cm}^{-2},
\label{eq:eq1-sec2-1}
\end{eqnarray}
\noindent where $R$ represents the radial coordinate in the midplane, $\Sigma_{\text{g}}^{0}$ and $\Sigma_{\text{p}}^{0}$ are normalization constants, and $\eta$ represents the discontinuity in the planetesimal surface density caused by the condensation of volatiles (particularly water) and it is given by
\begin{eqnarray}
  \eta= 
  \begin{cases}
    1 & \text{ if $R \ge R_{\text{ice}}$},  \\
    \\
    {\dfrac{1}{4}} & \text{ if $R < R_{\text{ice}}$}.
  \end{cases}
\label{eq:eq2-sec2-1}
\end{eqnarray}
$R_{\text{ice}}$, often called the snow line, corresponds to the radius where the temperature $T$ = 170 K. It is important to note here that the initial surface densities of realistic disks could not be represented by a simple power-law \citep{Andrews2010,Bitsch2015}.

In order to define the gas and planetesimal surface density profiles given by Eq.~\ref{eq:eq1-sec2-1}, it is necessary to specify the normalization constants $\Sigma_{\text{g}}^{0}$ and $\Sigma_{\text{p}}^{0}$. On the one hand, $\Sigma_{\text{g}}^{0}$ adopts the expression
\begin{eqnarray}
\Sigma_{\text{g}}^{0}= M_\text{d} / 2\pi (R_{\text{out}} - R_{\text{in}}),
\label{eq:nueva}
\end{eqnarray}
wherein $M_\text{d}$ is the mass of the disk, and $R_{\text{in}}$ and $R_{\text{out}}$ are the inner and outer radii of the disk, respectively. On the other hand, $\Sigma_{\text{p}}^{0}$ is given by $\Sigma_{\text{p}}^{0}= z_{\odot} 10^{[\text{Fe}/\text{H}]} \Sigma_{\text{g}}^{0}$, being $z_{\odot}= 0.0153$ the initial heavy element abundance of the Sun \citep{Lodders2009}, and $[\text{Fe}/\text{H}]$ the metallicity of the star\footnote{In the present work, we have assumed that the stellar content in heavy elements is a good measure of the overall abundance of heavy elements in the disk during formation time.}.

As we mentioned before, the aim of this work is to study the evolution of icy body reservoirs around low-mass stars in planetary scattering scenarios. In particular, we considered a low-mass star of $0.5~\text{M}_{\odot}$ and we assumed a temperature profile given by \citep{Ida2004}
\begin{eqnarray}
  \text{T}= 280 \left(\frac{R}{1~\text{au}} \right)^{-0.5} \left(\frac{L_{\star}}{\text{L}_{\odot}}\right)^{0.25}~\text{K},
  \label{eq:eq3-sec2-1}
\end{eqnarray}
\noindent where $L_{\star}$ represents the luminosity of the star. Then, according to \citet{Scalo2007}, the mass-luminosity relation for main sequence stars with masses less than or equal to the Sun's mass is obtained from the following expression
\begin{eqnarray}
  \log \left(\frac{L_{\star}}{\text{L}_{\odot}}\right) & = & 4.1 \log\left(\frac{M_{\star}}{\text{M}_{\odot}}\right)^3
  + 8.16 \log\left(\frac{M_{\star}}{\text{M}_{\odot}}\right)^2 + \nonumber \\
&& 7.11 \log\left(\frac{M_{\star}}{\text{M}_{\odot}}\right) + 0.065,
  \label{eq:eq4-sec2-1}
\end{eqnarray}
\noindent{where} $M_{\star}$ is the mass of the star. For a star of $0.5~\text{M}_{\odot}$, Eqs.~\ref{eq:eq3-sec2-1} and \ref{eq:eq4-sec2-1} imply that $R_{\text{ice}}$ is $\sim 0.5$~au.

\subsection{Model of disk evolution and planet formation}
\label{sec:sec2-2}

As we mentioned before, we want to find which parameters of the protoplanetary disk lead to the formation of three Jupiter-like planets located close to their limit of stability \citep{Marzari2014a}. To do this, we calculated the simultaneous in situ formation of three Jupiter-mass planets located at $\sim 0.5$ au, $\sim 0.8$ au, and $\sim 1.5$ au, using the model of planetary formation developed by \citet{Guilera2010} with some minor improvements.

We considered a disk between 0.1 au and 30 au using 5000 radial bins logarithmically equally spaced. The disk is characterized by a gaseous component and a population of planetesimals. In contrast to our previous work, we considered that the evolution of gaseous component is given by a viscous accretion disk \citep{Pringle1981} with photoevaporation \citep{Alexander2006,AlexanderArmitage2007}. Thus, the evolution of the gas surface density is governed by the equation
\begin{eqnarray}
  \frac{\partial \Sigma_g}{\partial t}= \frac{3}{R}\frac{\partial}{\partial R} \left[ R^{1/2} \frac{\partial}{\partial R} \left( \nu \Sigma_g R^{1/2}  \right) \right] + \dot{\Sigma}_w(R),
\label{eq:eq1-sec2-2}
\end{eqnarray}
where $\nu= \alpha c_s \text{H}_g$ is the viscosity \citep{ShakuraSunyaev1973}, being $\alpha$ a parameter, $c_s$ the sound speed, $\text{H}_g$ the scale height of the disk and $\dot{\Sigma}_w$ represents the sink term due to photoevaporation. On the other hand, the population of planetesimals obeys a continuity equation given by
\begin{eqnarray}
  \frac{\partial \Sigma_p}{\partial t} - \frac{1}{R}\frac{\partial}{\partial R} \bigg(Rv_{\text{mig}}(R)\Sigma_p\bigg) = \mathcal{F}(R),
\label{eq:eq2-sec2-2}
\end{eqnarray}
where $v_{\text{mig}}$ is the radial drift planetesimal velocity and $\mathcal{F}$ represents the sink term due to the accretion by the embryos\footnote{Here, we did not consider the planetesimal collisional evolution}. We considered that the evolution of the eccentricities and inclinations of the planetesimals is governed by two main processes: the gravitational stirring produced by the embryos, and the damping due to nebular gas drag. The gas drag also causes an orbital migration of the planetesimals. As in \citet{Guilera2014}, we assumed three different regimes: the Epstein regime, the Stokes regime and the quadratic regime.

Regarding the planets immersed in the disk, they grow by the accretion of planetesimals in the oligarchic regime \citep{Inaba2001}, and by the accretion of the surrounding gas solving the standard equations of stellar evolution for the envelope \citep[see][for a detailed explanation about our giant planet formation model]{Fortier2009,Guilera2010,Guilera2014}.

\subsection{Calculating initial conditions for N-body simulations}
\label{sec:sec2-3}

In the present work, we did not assume ad-hoc initial distributions of embryos and planetesimals for the N-body simulations. On the contrary, following the mechanism developed in \citet{deElia2013} and \citet{Ronco2015}, we applied the model of planetary formation performed by \citet{Guilera2010} with the goal of investigating which free parameters of the disk (i.e., mass of the disk, metallicity, planetesimal size, etc.) allow us to form three giant planets simultaneously in locations close to their stability limit. \citet{Ronco2015} found that initial conditions obtained from a planetary formation model can lead to different and more realistic accretion histories for the planets during the post-oligarchic growth with respect to ad-hoc initial conditions.

\begin{figure}[ht]
\centering
\includegraphics[scale= 0.35, angle= 270]{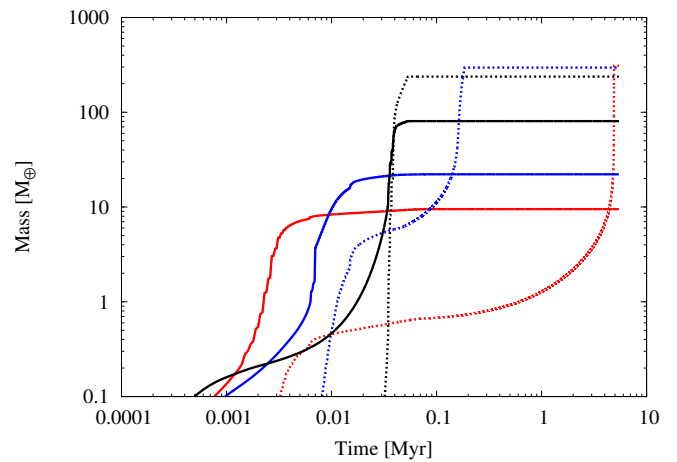}
\caption{
Time evolution of the mass of the cores (solid lines) and mass of the envelopes (dashed lines) for the three giant planets. Black lines
represent the outermost planet, blue lines the intermediate planet, and red line the innermost one.
}
\label{fig:fig1-sec2-3}
\end{figure}

\begin{figure}[ht]
\centering
\includegraphics[scale= 0.35, angle= 270]{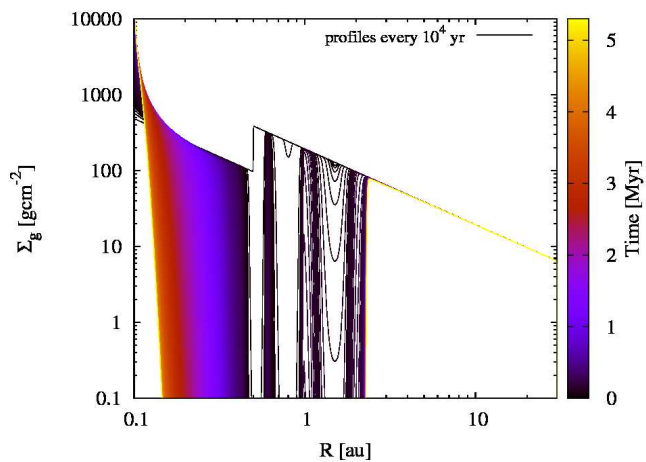}
\caption{
Time evolution of the radial profiles for the planetesimal surface density. As time advance, planets accrete the material in their feeding zones. In the inner part of the disk, planetesimal migration is efficient due to the high values of the gas density. However, in the outer part of the disk the profiles remain very similar respect to the initial one.
}
\label{fig:fig2-sec2-3}
\end{figure}

So, from a series of test simulations developed assuming different mass disk (between $0.01~\text{M}_{\odot}$ and $0.15~\text{M}_{\odot}$), planetesimal sizes (between 0.1 km and 100 km), and metallicities (between $[\text{Fe}/\text{H}]= -0.25$ and $[\text{Fe}/\text{H}]= 0.25$), we found that the particular combination of massive disks, planetesimals of $\sim 10$ km of radius, and high metallicities is the only one that allows us to form the three giant planets simultaneously. Since the main purpose of this work is to analyze the evolution of icy body reservoirs around low-mass stars resulting from planet-planet scattering events, we do not deepen on the details of these test simulations. However, it is important to mention that for low-mass and intermediate-mass disks, and for low-metallicity and intermediate-metallicity disks, there is not enough material to form massive cores and eventually gas giants. On the other hand, the test simulations with small planetesimals showed that the outermost planet in the disk acts as a barrier not allowing
the inner migration of the planetesimals, and thus the inner planets can not form massive cores. Finally, for large planetesimals of the order of 100 km, the formation timescales of the giant planets are greater than the dissipation timescale of the disk.

Figure~\ref{fig:fig1-sec2-3} shows the time evolution of the masses of the cores and envelopes for three giant planets. It is important to remark, that we did not assume limitation in the accretion of gas, so the growth of the planets is stopped when they reached one Jupiter mass. However, the planets continue exciting the planetesimal eccentricities and inclinations at a constant mass until the dissipation of the disk, which occurs at $\sim 5.3$~Myr, being consistent with the characteristic lifetime of observed protoplanetary disks \citep{Haisch2001,Mamajek2009,Pfalzner2014}. The results shown in Fig.~\ref{fig:fig1-sec2-3} correspond to a massive disk of $M_{\text d}= 0.15~\text{M}_{\odot}$, with a high metallicity of $[\text{Fe}/\text{H}]= 0.25$, using a value of $\alpha= 10^{-4}$, and planetesimals of 10 km of radius. It is important remark that massive disks could be present in a low-mass star. In fact, \citet{Andrews2010} estimated a mass of 0.143 M$_{\odot}$ for the disk GSS 39, which is associated to a M0-type star of 0.6 M$_{\odot}$.

We note that we did not take into account type I and type II migrations in our simulations. Type I migration predicts rapid inward migration rates in an idealized isothermal disk \citep{Tanaka2002}. However, studies in more realistic disks show that type I migration rates could substantially change \citep{Paardekooper2010,Paardekooper2011}. More recently, \citet{Benitez-LLambay2015} found that if the released energy by the planet due to accretion of solid material is taken into account, type I migration can be significantly slow down, cancel, and even reverse. On the other hand, type II migration is likely to be an important effect in multi-planet systems. However, calculations of type II migration rates in such systems result to be a complex task, and a simplistic single-planet treatment could lead to different results. But, for very small values of the alpha-viscosity parameter ($\alpha<10^{-5}$), type II migration rates could became negligible. Such values could be possible to achieve in the often called dead zones \citep{Matsumura2009}.

In Fig.~\ref{fig:fig2-sec2-3}, the time evolution of the radial profile of the planetesimal surface density is shown. We can see that planetesimals are quickly accreted by the planet. However, behind the outermost giant planet, planetesimals practically keep the same structure. This is because for planetesimals of $\sim 10$~km of radius, the inward drift is negligible.

\subsection{N-body simulations: Characterization, parameters, and initial conditions}

To study the dynamical evolution of our systems, we carried out our N-body simulations using the Mercury code, which was developed by \citet{Chambers1999}. In particular, we used the RA15 version of the RADAU numerical integrator with an accuracy parameter of $10^{-12}$ \citep{Everhart1985}. In agreement with \citet{Marzari2014b}, we decided to make use of such an integrator due to its stability and precision when dealing with gravitational encounters with Jupiter-mass bodies.

Here, we studied the orbital evolution of three Jupiter-mass planets close to their stability limit together with an outer planetesimal disk around M-type stars of 0.5 M$_{\odot}$. In order to make use of the N-body code, it is necessary to specify physical and orbital parameters for the gas giant planets and the planetesimals.

As for the giants, all our simulations were developed assuming three Jupiter-mass planets with a physical density of 1.33 g cm$^{-3}$. We assigned to the inner gas giant a semimajor axis $a_{1}$ of 0.51 au in all simulations, which matches the location of the snow line around a 0.5 M$_{\odot}$ star, such as was mentioned in Sect. 2.1. Then, the other two Jupiter-mass planets were located on orbits whose semimajor axes had values close to their stability limit in agreement with \citet{Marzari2014a}. In fact, the second and third giant planet of the system adopted semimajor axes $a_{2}$ and $a_{3}$ given by $a_{2}$ = $a_{1}$ + $K_{1}$ $R_{H}^{1,2}$ and $a_{3}$ = $a_{2}$ + $K_{2}$ $R_{H}^{2,3}$, where $R_{H}^{1,2}$ and $R_{H}^{2,3}$ represent the mutual Hill's spheres between the consecutive planets 1-2 and 2-3, respectively. We note that we adopted the two dimensional approach in terms of $K$ parameter to find the transition between stable and unstable systems that may lead to planet-planet scattering, such as \citet{Marzari2014a} suggested. According to \citet{Marzari2014b}, $K_{1}$ and $K_{2}$ parameters were selected randomly from a uniform distribution between 2.8 and 3.4 and between 4.0 and 5.8, respectively. These ranges adopted for $K_{1}$ and $K_{2}$ parameters were chosen by the need of simulating systems that become dynamically unstable on a short timescale due to computational reasons. It is important to mention that the convergent migration in gaseous protoplanetary disks could lead to mean motion resonances (MMRs) between giant planets \citep{Snellgrove2001, LeePeale2002}. In fact, models show that the capture in the 2:1 and 3:2 MMRs is a process of common occurrence \citep{Thommes2005,PierensNelson2008,LeeThommesRasio2008}. However, turbulence driven by magneto-rotational instabilities can act to remove planets from resonances. In fact, \citet{Adams2008} estimated that the fraction of resonant systems that survive over a disk lifetime of 1 Myr is of order 0.01. From such considerations, the three Jupiter-mass planets of our simulations were not initially in mutual MMRs but they were located very close to their stability limit derived by \citet{Marzari2014a}. This criterion was also adopted by several previous works about planet-planet scattering such as \citet{Raymond2008}, \citet{Raymond2009a}, \citet{Raymond2009b}, \citet{Raymond2010}, \citet{Raymond2013}, and \citet{Marzari2014b}. We want to remark that the main internal and external MMRs (2:1, 3:1, 3:2, 4:3, 5:3) fall between the initial locations of the giant planets in our systems of work. On the other hand, eccentricities and inclinations of the giant planets were taken randomly considering values lower than 0.01 and 0.5$^{\circ}$ respectively. Moreover, the mean anomaly $M$, the longitude of ascending node $\Omega$, and the argument of periapse $\omega$ were randomly assigned between 0$^{\circ}$ and 360$^{\circ}$. It is important to remark that these initial orbital parameters assigned to the gaseous giants were referred to the midplane of the original protoplanetary disk.

To model the planetesimal population, we assumed an outer disk composed by 1000 massless bodies up to 30 au. In each simulation, the inner edge of the belt was located at four Hill's radii of the outermost Jupiter-mass planet of the system. The semimajor axis of each test particle was assigned from the surface density profile derived in Sect. 2.3 making use of the semi-analytical model. According to that analysis, the planetesimals followed an $R^{-1}$ surface density profile, so the number of the test particles was distributed uniformly across the disk. Regarding the other orbital parameters, eccentricities and inclinations were selected randomly from a uniform distribution assuming values lower than 0.001 and 0.5$^{\circ}$ respectively, while the mean anomaly $M$, the longitude of ascending node $\Omega$, and the argument of periapse $\omega$ were randomly assigned between 0$^{\circ}$ and 360$^{\circ}$. As for the giant planets, these initial orbital parameters assigned to the planetesimals were referred to the midplane of the original protoplanetary disk.

It is worth noting that our N-body simulations were developed assuming an M-type star of 0.5 M$_{\odot}$ with a non-realistic size of 0.03 au of radius. Thus, all those bodies with perihelion distances less than 0.03 au were removed from the system assuming a collision with the central star. Moreover, particles were considered to be ejected if they reached distances greater than 1000 au from the central star.

\section{Results}

In this section, we describe the results of N-body simulations aimed at studying the final planetary architecture and the formation and evolution of reservoirs of icy small bodies. In particular, these simulations were carried out in systems around  low-mass stars of 0.5 M$_{\sun}$ in scattering planetary scenarios produced by strong dynamical instabilities.

We carried out a total of 80 N-body simulations, each of which started with three Jupiter-mass planets close their stability limit and an outer small body reservoir. The initial configuration proposed for the gaseous giants led to strong dynamical instability events in the systems of study. Such events produced collisions between planets, ejections of one or more giants of the system, as well as collisions with the central star. 

In 50 of 80 numerical simulations, the planet-planet scattering event led to collisions between two Jupiter-mass gaseous giants. The N-body code used here treats collisions as inelastic mergers conserving linear momentum. However, we consider that assuming perfect mergers between Jupiter-mass planets as a result of a collision is not an appropriate hypothesis. In fact, using hydrodynamical simulations, \citet{Inamdar2016} calculated the envelope fraction lost due to giant impacts for initial envelope fractions of 1\% - 10\%. Even though those authors analyzed only planets with masses in the range of the Neptunes and the so-called super-Earths, they obtained that a single collision between similarly sized planets can easily reduce the envelope-to-core-mass ratio by a factor of two. From these considerations, our analysis does not consider those simulations in which collisions between gaseous giants are produced because the main physical and dynamical effects that happen in such impact events are not clearly understood.

In 30 of 80 numerical simulations, one, two, or three Jupiter-mass planets were removed from the system after the dynamical instability event. From the 30 simulations carried out, in 21 of them (70\%) only one single giant planet survived, two giant planets survived in 6 (20\%), and none planet survived in the other three runs (10\%). Interestingly, we found small bodies reservoirs in the three cases. It is worth remarking that the outer test particles showed very different dynamical features depending on the number of Jupiter-mass planets surviving in the system. From this, we decided to focus the present research on the systems in which a single giant planet survived after the dynamical instability event together with an outer small body reservoir\footnote{All simulations presented here maintained excellent energy conservation properties ($dE/E$ $<$ 6 $\times$ 10$^{-11}$).}. An study concerning the dynamical evolution of outer minor planet populations in systems with two surviving Jupiter-mass planets will be presented in a forthcoming paper.

An analysis of our results indicates that the planetary scattering events lead to the formation of reservoirs of small bodies whose
representative members show different dynamical behaviors. From our numerical simulations, we distinguish three different types of test particles:
\begin{itemize}
\item[$*$] {\bf Type-P:} Particles that have prograde orbits along their evolution, ie. they keep their inclinations $<$ 90$\degr$.
\item[$*$] {\bf Type-R:} Particles that have retrograde orbits along their evolution, ie. they keep their inclination $>$ 90$\degr$.
\item[$*$] {\bf Type-F:} Particles whose orbital plane flips from prograde to retrograde along their evolution (or vice versa).
\end{itemize}

An detailed description of the results obtained in our set of N-body simulations is presented in the next subsections. All N-body simulations analyzed here were integrated for a time span of 100 Myr. Moreover, in order to analyze properly our results, we considered the orbital parameters of test particles and the surviving planet are referenced to the barycenter and invariant plane of the system (star + planet), whose x-axis was chosen to coincide with the surviving planet's periapse.

\subsection{Planetary configuration}

\begin{figure}
\centering
\includegraphics[scale= 0.35, angle= 270]{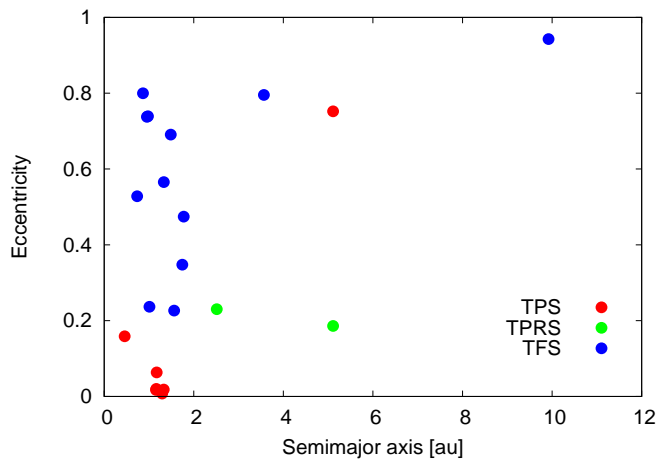}
\caption{
Orbital parameters $e$ vs. $a$ of the surviving planets. The red circles represent planets resulting in systems with Type-P test particles, that we called TPS. The green ones illustrate planets belonging to
systems with Type-P and Type-R particles, which we called TPRS. Finally, the blue circles represent planets in systems composed of at least one Type-F test particle, whose abbreviation is TFS.
}
\label{fig:fig3}
\end{figure}

Figure~\ref{fig:fig3} shows the final distribution of the surviving Jupiter-mass planets in the orbital parameter space eccentricity vs. semimajor axis. The final values associated with the semimajor axis and eccentricity of such planets range from 0.5 au to 10 au and 0.01 to 0.94, respectively. The planets surviving in systems composed only of Type-P test particles are indicated in the figure, as are planets associated to systems where Type-P particles coexist with Type-R particles. We remark that our numerical simulations produced only two systems in which the outer small body reservoirs are composed of Type-P and -R particles. However, it is worth mentioning that these systems have 588 (744) Type-P test particles and only 2 (1) Type-R particles. Finally, planets surviving in systems where at least one Type-F particle exists in the small body population are also plotted. The systems composed of Jupiter-mass planets represented for red, green, and blue circles will be called TPS, TPRS, and TFS, respectively.

As the reader can see in Fig.~\ref{fig:fig3}, all systems that have at least one Type-F particle are dominated by a surviving planet with an eccentricity $e >$ 0.2. However, it is worth noting that the presence of a Jupiter-mass planet with an eccentricity $e >$ 0.2 does not imply the existence of Type-F particles in the system. In fact, the planet with $a$ = 5.1 au, $e$ = 0.75 belongs to a system whose small body reservoir is composed only of Type-P particles. A very interesting particular case obtained from our simulations is given by a system composed only of Type-F test particles. In such a system, the Jupiter-mass planet shows a semimajor axis and an eccentricity of 9.9 au and 0.94, respectively.

\begin{figure}
\centering
\includegraphics[scale= 0.35, angle= 270]{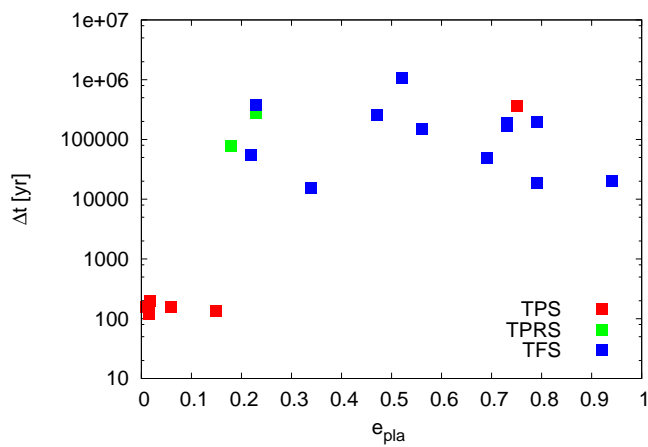}
\caption{
Duration of instability events vs. eccentricity of the surviving planet.
The red squares represent the resulting systems with only Type-P test particles (TPS), the green squares represent the resulting systems with Type-P and Type-R particles (TPRS), and the blue ones represent the resulting systems with at least one Type-F particle (TFS).
}
\label{fig:fig4}
\end{figure}

We also analyzed the dynamical instability timescale associated to our resulting systems. The duration of the instability event is defined as the period between the first close encounter between the giant planets and the time in which a second planet is removed, leading to a system with a single Jupiter-mass planet as survivor. Figure~\ref{fig:fig4} shows the duration of the dynamical instability ($\Delta$t) as a function of the eccentricity of the surviving planet ($e_{\text{pla}}$). The resulting systems with at least one Type-F particle (TFS) and the resulting systems with only Type-P test particles are both indicated on the figure. In general terms, the TPSs show $\Delta$t less than those associated to TFSs. Indeed, in the cases where $\Delta$t is short, the material is not dispersed, leaving a single small body reservoir formed by the Type-P particles.

\subsection{Resulting ice body reservoirs: general analysis}

\begin{figure*}
\includegraphics[angle=270, width=0.95\textwidth]{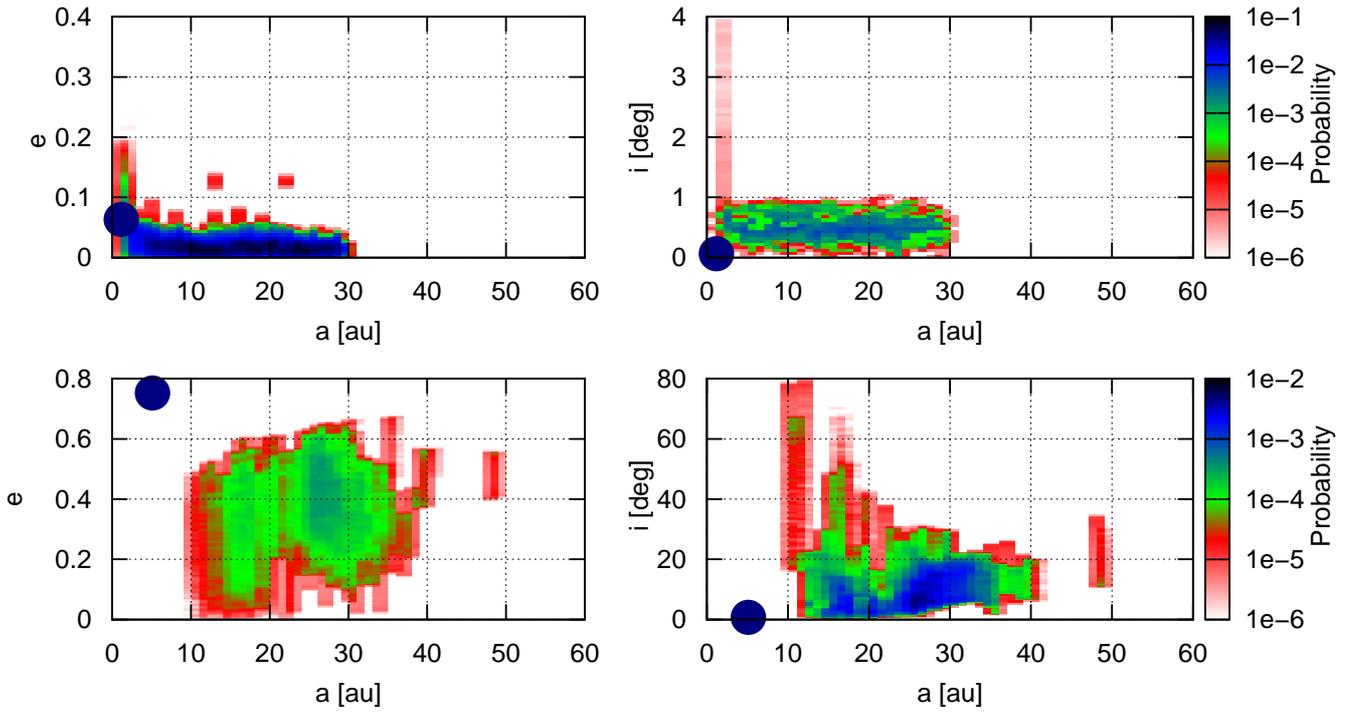}
\caption{
Occupation $e$ vs. $a$ and $i$ vs. $a$ of systems composed of Type-P particles. The color zones of these maps are regions
with different degrees of probability where the particles can be found. The blue filled circle represents the planet surviving in each
system. Top panels: reservoir of small bodies with low eccentricities ($e$ $<$ 0.05) and low inclinations ($i$ $<$ 1$\degr$). The
surviving Jupiter-mass planet has values of semimajor axis and eccentricity of 1.16 au and 0.06, respectively. Bottom panels: population
of small bodies, which reach eccentricities $e$ $<$ 0.6 and inclinations $i$ $<$ 80$\degr$. The surviving Jupiter-mass planet shows a
semimajor axis $a =$ 5.1 au and an eccentricity $e =$ 0.75.
}
\label{fig:fig5}
\end{figure*}

\begin{figure*}
\includegraphics[angle=270, width=0.95\textwidth]{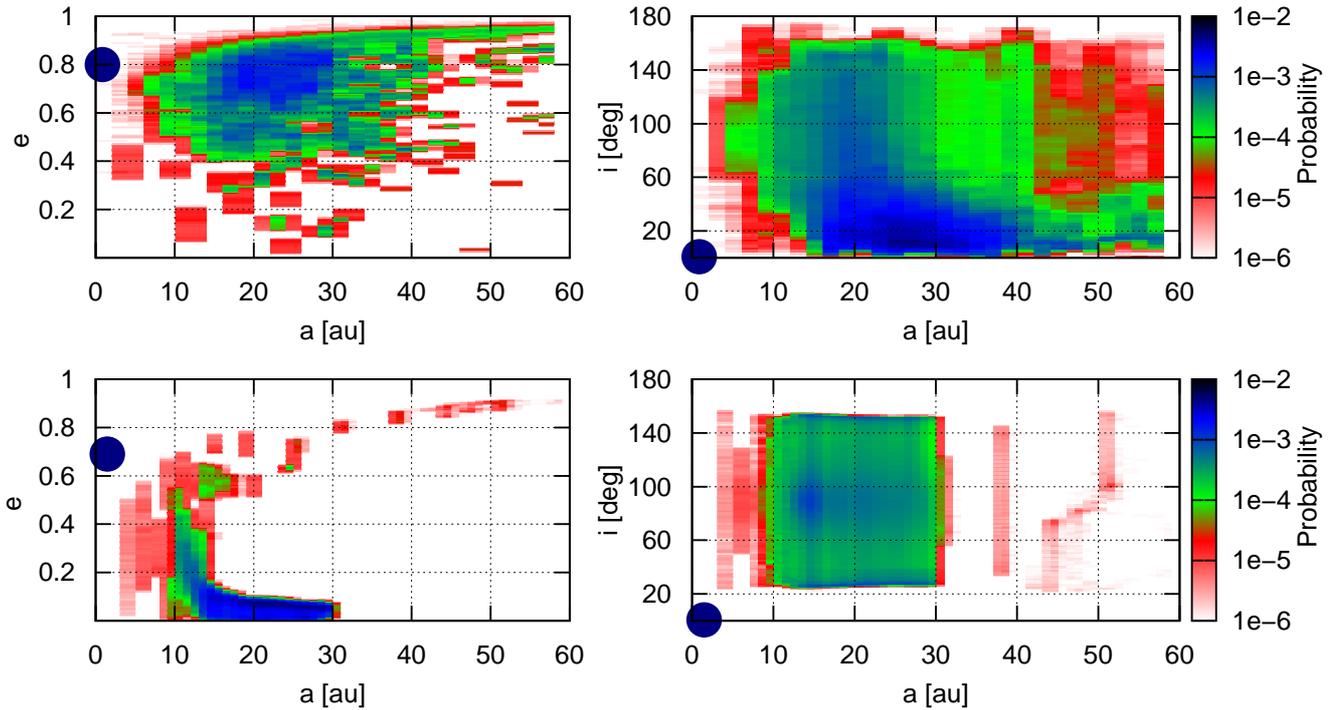}
\caption{Occupation $e$ vs. $a$ and $i$ vs. $a$ of systems with Type-F particles.
The color zones of these maps are regions with different degrees of probability where the particles can be found. The blue filled circle represents the planet surviving in each system. Top panels: system with three different small body populations, which are composed of Type-P, Type-R and Type-F particles. In general terms, the particles have from moderate to high eccentricities and a wide range inclinations. The surviving Jupiter-mass planet has values of semimajor axis and eccentricity of 0.86 au and 0.8, respectively. Bottom panels: system with two different small body reservoirs, which are composed of Type-P and Type-F particles. The most notable feature is that the Type-F particles show very low eccentricities ($e$ $\lesssim$ 0.2). The surviving Jupiter-mass planet shows a semimajor axis $a =$ 1.48 au and an eccentricity $e =$ 0.69.
}
\label{fig:fig6}
\end{figure*}

\begin{figure*}
\centering
\includegraphics[angle=0, width=0.45\textwidth]{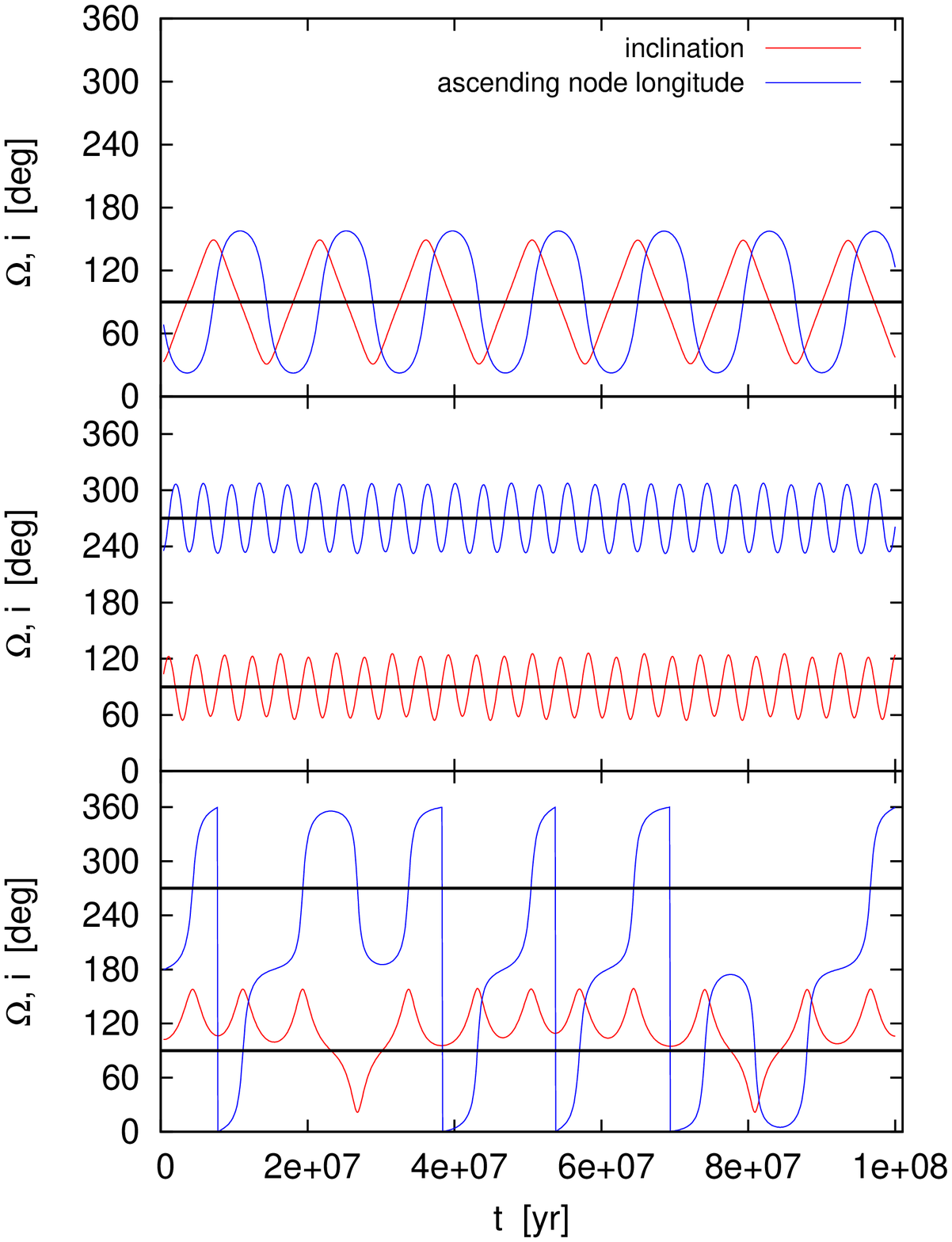}
\includegraphics[angle=0, width=0.45\textwidth]{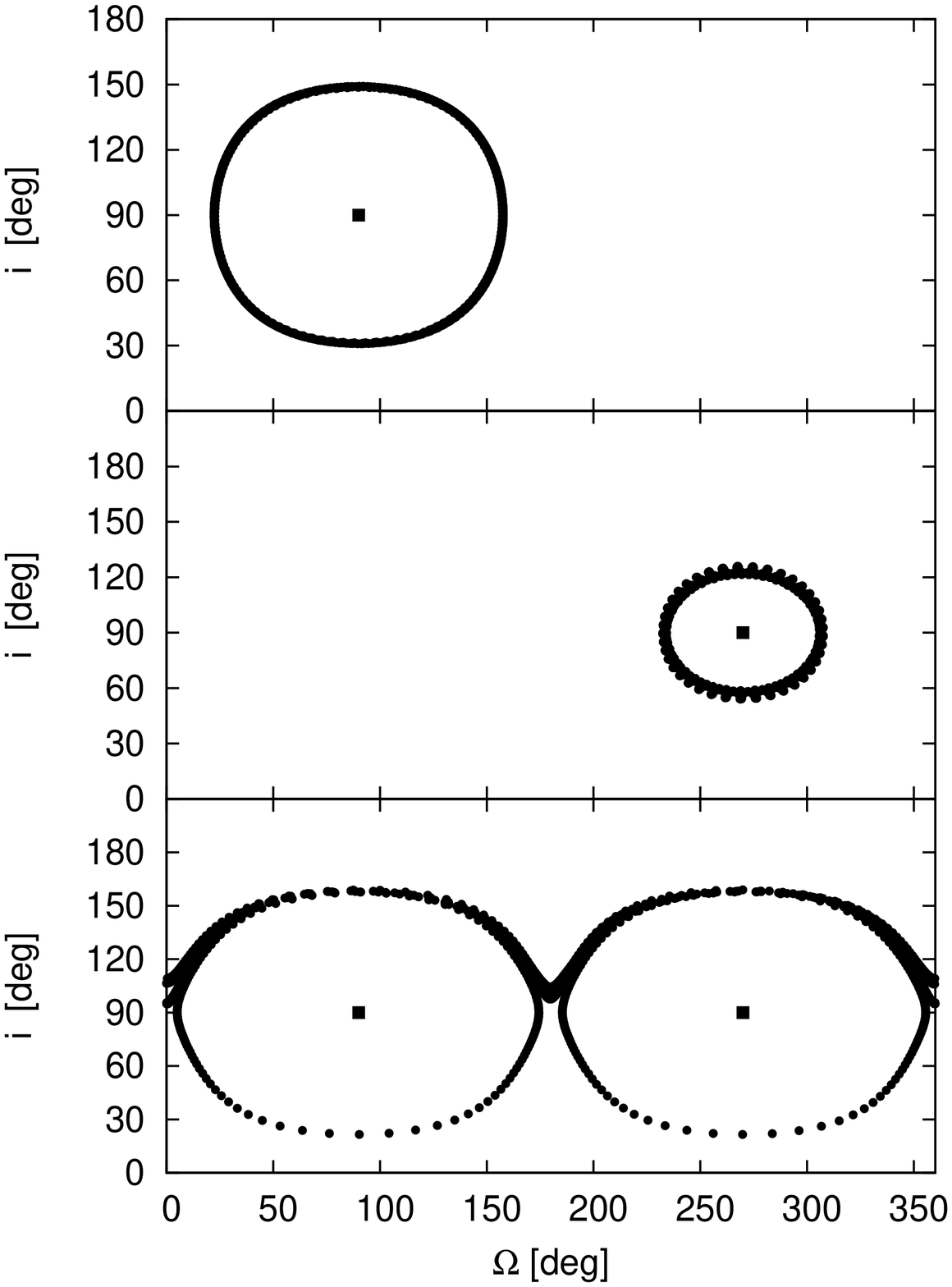}
\caption{
Evolution in time of the inclination $i$ (red line) and the ascending node longitude $\Omega$ (blue line) of three different test particles
(left side) of the same simulation. On the right side, the coupling of inclination $i$ and the ascending node longitude $\Omega$ for the same
particles. The semimajor axis and eccentricity of the surviving giant planet are $a =$ 0.95 ua and $e =$ 0.73.
Top panels: the particle's orbit flips while the $\Omega$ librates around 90$\degr$ (black line). The orbital parameters immediately after the instability event (hereafter, post IE parameters) for this particle are $a_{1}$ $=$ 16.9 ua, $e_{1}$ $=$ 0.2, $i_{1}$ $=$ 31.4$\degr$, $\omega_{1}$ $=$ 131.7$\degr$, and $\Omega_{1}$ $=$ 79$\degr$. Middle panels: the particle's orbit flips while the $\Omega$ librates around 270$\degr$ (black line). The post IE orbital parameters for this particle are $a_{2}$ $=$ 16.61 ua, $e_{2}$ $=$ 0.54, $i_{2}$ $=$ 93.2$\degr$, $\omega_{2}$ $=$ 2$\degr$, and $\Omega_{2}$ $=$ 232.5$\degr$. Bottom panels: the orbit flips while the libration center of $\Omega$ changes from 270$\degr$ to 90$\degr$ along its evolution (black lines). The post IE orbital parameters for this particle are $a_{3}$ $=$ 23.34 ua, $e_{3}$ $=$ 0.77, $i_{3}$ $=$ 102.4$\degr$, $\omega_{3}$ $=$ 53.19$\degr$, and $\Omega_{3}$ $=$ 178.8$\degr$.
}
\label{fig:fig7}
\end{figure*}

From our study, seven of the 21 simulations ended with a single population of small bodies whose members are Type-P particles. Representative systems of this class are shown in Fig.~\ref{fig:fig5}, which illustrates occupation maps in the orbital planes ($a$,$e$) and ($a$,$i$). Each plot shows the normalized time fraction spent by the test particles in different regions of the ($a$,$e$) and ($a$,$i$) planes during the whole integration time. The color code is indicative of the portion of time or permanence time spent in each zone being blue for most visited regions and red for those least visited. Those plots represent also probability maps since the time spent in each zone is normalized. In general terms, six of these seven systems show similar evolutionary history associated to giant planets and the reservoirs of small bodies. A common feature in these systems is that the instability event is produced very quickly.
In such cases, one of the planets was immersed into the planetesimal disk by 10$^{2}$-10$^{3}$~yr after which it was removed from the system, keeping more than 99$\%$ of the original particles. The systems aforementioned are composed by a giant planet with $e$ $<$ 0.15 and a compact small body reservoir as a cold disk (top panels of Fig.~\ref{fig:fig5}). In general terms, these cold disks have semimajor axes 2~au $<$ $a$ $<$ 30~au, eccentricities $e$ $<$ 0.1, and inclinations $i$ $<$ 1$\degr$. A significant result that we obtained is that this class of systems preserves the dynamics properties of the original planetesimal disk.

The remaining system, which has only Type-P particles as reservoirs of small bodies, shows a different evolutionary history in
comparison with the other six. Such a system is composed by a giant planet with $a$ $=$ 5.1 au and $e$ $=$ 0.75, and a population of minor
planets, which have $e$ $<$ 0.6 and $i$ $<$ 60$\degr$. The bottom panels of Fig.~\ref{fig:fig5} show occupation maps of this system. The discrepancy
between this system and those described in the previous paragraph is due to the combination of two factors. On the one hand, the dynamical
instability time-scale is significantly longer, having an estimated duration 3.5 $\times$ 10$^{5}$ $\mathrm {yr}$. On the other hand, the
surviving planet has a highly eccentric orbit. In fact, both effects produce the effective removal of particles from the system, keeping
only 30$\%$ of the initial population, and the excitation of the planetesimal disk.

12 of 21 simulations have at least one Type-F particle in the system. In most of these simulations, Type-P, Type-R, and Type-F particles
coexist in the system producing three different small body reservoirs. Occupation maps of two representative planetary systems with such
features are shown in Fig.~\ref{fig:fig6}. In particular, a system composed of Type-P, Type-R, and Type-F particles with moderate and high
eccentricities is shown in the top panels of Fig.~\ref{fig:fig6}. In this case, the surviving giant planet shows values of the semimajor axis and
eccentricity of 0.86 au and 0.8, respectively. On the other hand, a very interesting system is shown in the bottom panels of Fig.~\ref{fig:fig6}. In
such a system, Type-P and Type-F particles coexist. The most notable feature observed in that system is that the particle's orbits flip
for very low values of the eccentricity ($e <$ 0.2). In this case, the surviving Jupiter-mass planet has values of the semimajor axis and
eccentricity of 1.48 au and 0.69, respectively.

\subsection{Analysis of Type-F particles}

A relevant result is that all Type-F particles of our simulations show an oscillation of the ascending node longitude $\Omega$. It is worth remembering that $\Omega$ of the test particles is measured relative to the surviving planet's periapse in agreement with \citet{Lithwick2011}. When $\Omega$ oscillates, the orbital plane of the test particle rotates around of the nodal line, and from this, the particle's orbit flips from prograde to retrograde and back again. This result is consistent with that derived by \citet{Naoz2017}. These authors have recently studied the secular evolution of an outer test particle under the effects of an inner eccentric planet up to the octupole level of approximation. \citet{Naoz2017} showed that in the quadrupole level of approximation, the ascending node longitude of the test particle $\Omega$ has two different classes of trajectories: libration and circulation. In particular, the libration mode of $\Omega$ leads to oscillations of the test particle's inclination $i$ around $i = 90\degr$ by producing the flipping of the orbit. 

Our study suggests that the flipping of an orbit is correlated with oscillations of $\Omega$ around 90$\degr$ or/and 270$\degr$, which is in agreement with \citet{Naoz2017}. Figure~\ref{fig:fig7} shows the dynamics of three different Type-F particles of a same simulation. The evolution in time of the inclination $i$ and the ascending node longitude $\Omega$ for the three particles can be seen in the left side of Fig.~\ref{fig:fig7}, while the coupling of $i$ and $\Omega$ is shown on the right side of such figure. In the top and middle panels, the libration center of $\Omega$ is around 90$\degr$ and 270$\degr$, respectively. In the bottom panels, the ascending node longitude $\Omega$ of the test particle circulates and librates along its evolution. In particular, as $\Omega$ oscillates, its libration center is around 90$\degr$ or 270$\degr$. This result is consistent with that obtained by \citet{Naoz2017}, who showed that introducing the octupole level of secular approximation allows for transition between the libration and circulation modes. This behavior is observed in a few Type-F particles of our simulations.

\begin{figure}[t]
\centering
\includegraphics[angle=0, width=0.5\textwidth]{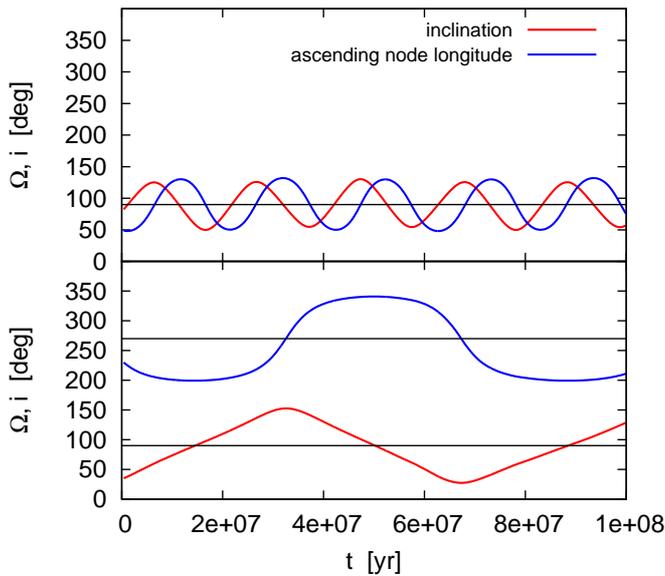}
\caption{
Evolution in time of the inclination $i$ and the ascending node longitude $\Omega$ for two different Type-F particles of the same simulation.
The semimajor axis and eccentricity of the surviving giant planet are $a =$ 0.95 ua and $e =$ 0.73, while the post IE orbital parameters of the test particles are $a_{1}$ $=$ 47.84 ua, $e_{1}$ $=$ 0.86, $i_{1}$ $=$ 80.2$\degr$, $\omega_{1}$ $=$ 178.37$\degr$, $\Omega_{1}$ $=$ 49.2$\degr$ (top panel), and $a_{2}$ $=$ 124.95 ua, $e_{2}$ $=$ 0.96, $i_{2}$ $=$ 34.27$\degr$, $\omega_{2}$ $=$ 285$\degr$, and $\Omega_{2}$ $=$ 231.79$\degr$ (bottom panel). It is worth noting that the libration periods associated to $i$ and $\Omega$ are equivalent and the oscillations of both angles are out of phase by a quarter period.
}
\label{fig:fig9}
\end{figure}

\begin{figure*}
\centering
\includegraphics[angle=0, width=0.45 \textwidth]{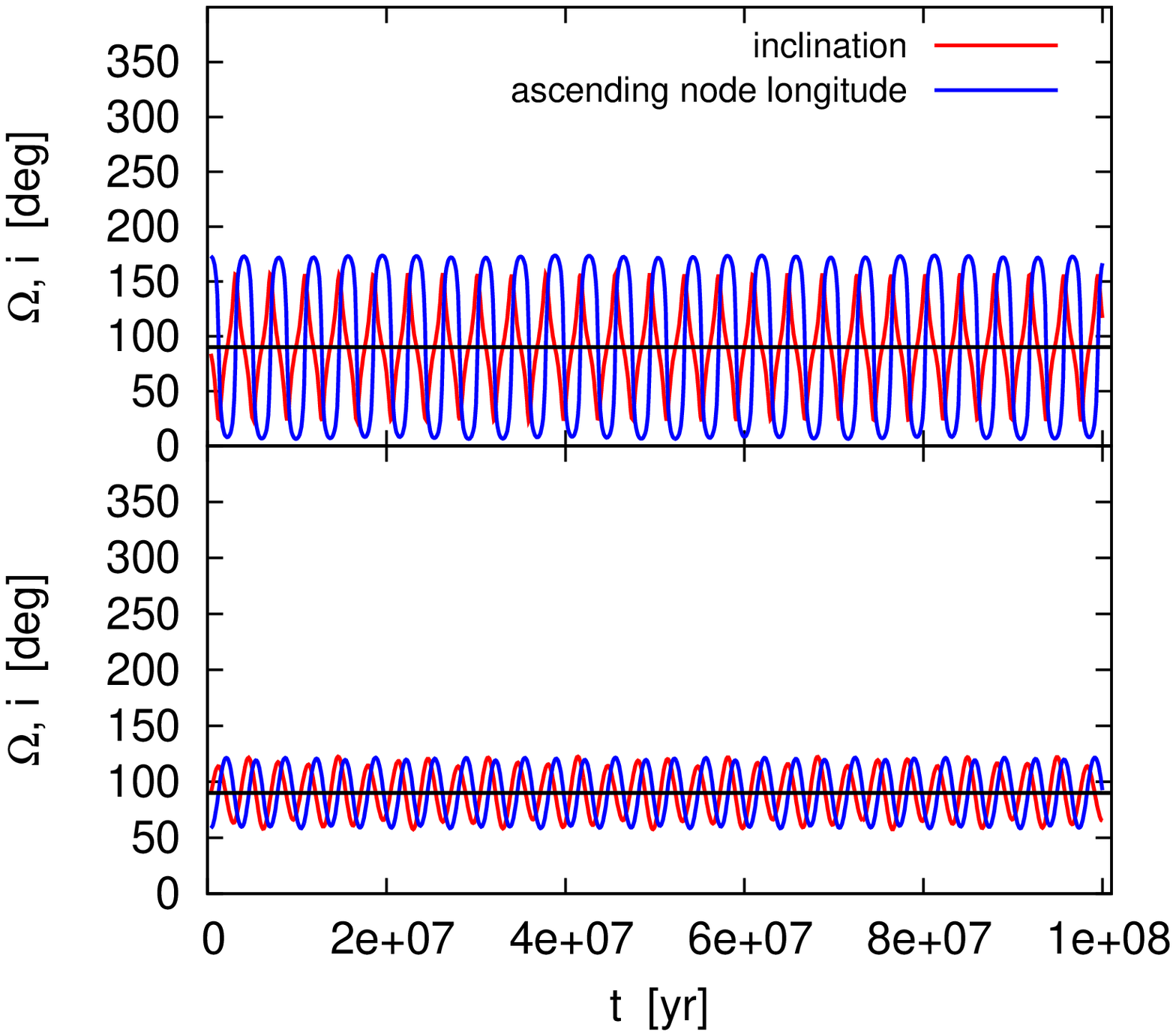}
\includegraphics[angle=0, width=0.45 \textwidth]{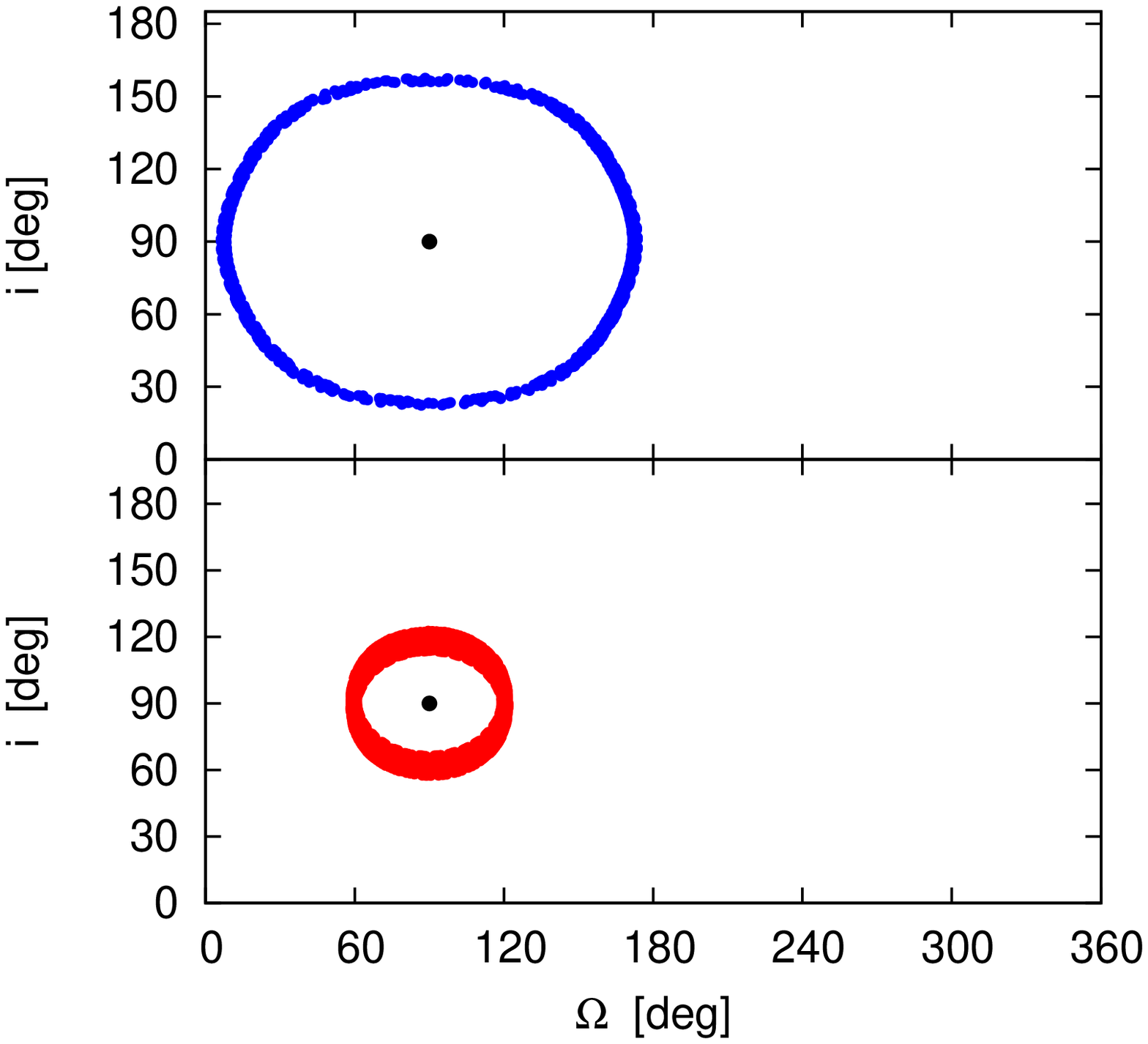}
\caption{
Evolution in time of the inclination $i$ (red line) and the ascending node longitude $\Omega$ (blue line) of two different test particles
(left side) of the same simulation. On the right side, the coupling of inclination $i$ and the ascending node longitude $\Omega$ for the same
particles. The red and blue dots show the values of $i$ and $\Omega$ for particles flipping with small and large amplitudes, respectively.
The black circle represent the center of libration for both angles. The larger the libration amplitude of $\Omega$, the larger the libration
amplitude of $i$. The semimajor axis and eccentricity of the surviving giant planet are $a =$ 0.95 ua and $e =$ 0.73.
The top panels are represented by a test particle whose post IE orbital parameters are $a_{1}$ $=$ 11.6 ua, $e_{1}$ $=$ 0.48, $i_{1}$ $=$ 93$\degr$, $\omega_{1}$ $=$ 9.22$\degr$, and $\Omega_{1}$ $=$ 173.48$\degr$. The post IE orbital parameters for the test particle in the bottom panels are $a_{2}$ $=$ 21.35 ua, $e_{2}$ $=$ 0.74, $i_{2}$ $=$ 80.3$\degr$, $\omega_{2}$ $=$ 167.9$\degr$, and $\Omega_{2}$ $=$ 59.7$\degr$.
}
\label{fig:fig8}
\end{figure*}

We want highlight that the oscillation of $\Omega$ is a necessary and sufficient condition for that the flipping mechanism can be activated.
When the particle's orbit is prograde or retrograde during its evolution, $\Omega$ circulates, while when the particle's orbit flips, $\Omega$ oscillates. This behavior, which occurs for all Type-F particles of our simulations, can be clearly seen in the bottom panels of Fig.~\ref{fig:fig7}. The test particle begins with a retrograde orbit and $\Omega$ circulates. At 20 Myr the particle's orbit flips and $\Omega$ oscillates around 270$\degr$. At 35 Myr the orbit is retrograde and $\Omega$ circulates again. Finally, at $\sim$ 75 Myr, the particle's orbit flips and $\Omega$ oscillates around 90$\degr$.

We observed strong correlations between the ascending node longitude $\Omega$ and the inclination $i$ of the Type-F particles. On the one hand, our results suggest that the libration periods of $\Omega$ and $i$ are equal for each Type-F particle obtained in our simulations. Another interesting result of our study indicates that the temporal evolution of $\Omega$ and $i$ are out of phase by a quarter period for all Type-F particles. This property can be seen in Fig.~\ref{fig:fig9}, which represents the evolution in time of $\Omega$ and $i$ for two different Type-F particles associated to a same simulation. These results are in agreement with \citet{Naoz2017}, who found that the extreme values of $i$ for the libration mode are located at $\Omega = 90\degr$ or $270\degr$, while $\Omega$ reaches minimum and maximum values at $i = 90\degr$. Moreover, as the reader can see in both panels of Fig.~\ref{fig:fig9}, when the particle's orbit changes from prograde (retrograde) to retrograde (prograde), $\Omega$ reaches a minimum (maximum) value. This is consistent with the results derived by \citet{Naoz2017}, who showed that $\text{d}\Omega/\text{d}{i} \propto -\cos i\sin i$ \citep[see Eq. 17 from][]{Naoz2017}.

Our study also suggests a strong correlation between the libration amplitudes of $\Omega$ and $i$ of all Type-F particles obtained in our simulations. In fact, the larger the libration amplitude of $\Omega$, the larger the libration amplitude of $i$. This is according to \citet{Naoz2017}, who showed that $\cos 2\Omega_{\text{min}} \propto -\sin^{2} i_{\text{min}}$, where $i_{\text{min}}$ and $\Omega_{\text{min}}$ represent the minimum values of the inclination and the ascending node longitude of the test particle, respectively \citep[see Eq. 21 from][]{Naoz2017}. This correlation can be seen in Fig.~\ref{fig:fig8} for the case of two different Type-F particles of a same simulation.

In general terms, when the ascending node longitude $\Omega$ and the inclination $i$ librate, the periapse argument $\omega$ circulates.
However, it is worth mentioning that librations of $\Omega$, $i$, and $\omega$ have been also observed in a few test particles of our simulations.

\begin{figure}[ht]
\centering
\includegraphics[angle=0, width=0.5\textwidth]{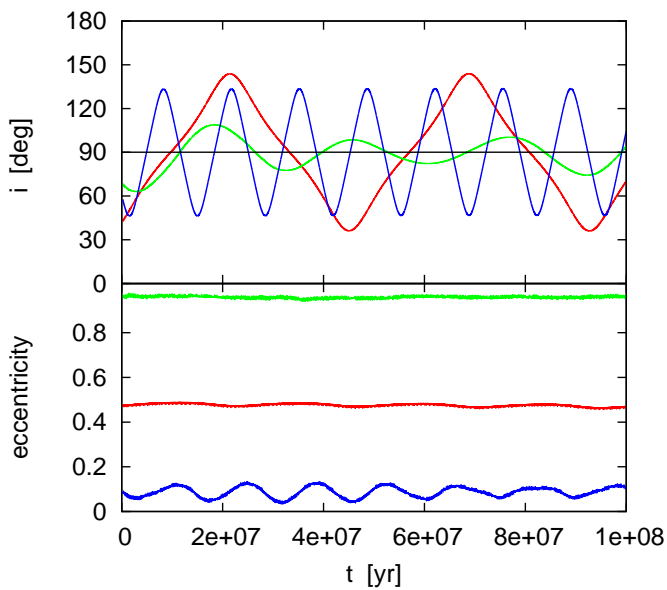}
\caption{
Evolution in time of the inclination $i$ (top panel) and the eccentricity $e$ (bottom panel) for three different Type-F particles corresponding
to the same simulation. The semimajor axis and eccentricity of the surviving planet are $a =$ 1.33 ua and $e =$ 0.56.
For the case of the test particles, the post IE orbital parameters are $a_{1}$ $=$ 19.14 ua, $e_{1}$ $=$ 0.08, $i_{1}$ $=$ 57.9$\degr$, 
$\omega_{1}$ $=$ 304.28$\degr$, $\Omega_{1}$ $=$ 308.23$\degr$ (blue line); $a_{2}$ $=$ 27 ua, $e_{2}$ $=$ 0.47, $i_{2}$ $=$ 42.7$\degr$, 
$\omega_{2}$ $=$ 110.25$\degr$, $\Omega_{2}$ $=$ 234.15$\degr$ (red line); $a_{3}$ $=$ 109.38 ua, $e_{3}$ $=$ 0.96, $i_{3}$ $=$ 68.2$\degr$,
$\omega_{3}$ $=$ 355$\degr$, $\Omega_{3}$ $=$ 282.2$\degr$ (green line).
}
\label{fig:fig10}
\end{figure}

A very interesting result derived from our analysis suggests that the particle's orbit can flip for any value of its eccentricity $e$.
Figure~\ref{fig:fig10} shows the evolution of the inclination $i$ and the eccentricity $e$ for three different Type-F particles of a same 
simulation with a wide diversity of eccentricities. In fact, our simulations show Type-F particles with quasi-circular orbits, with moderated 
eccentricities, as well as with very high eccentricities close to unity (Fig.~\ref{fig:fig10}).

\begin{figure}[ht]
\centering
\includegraphics[angle=0, width=0.5\textwidth]{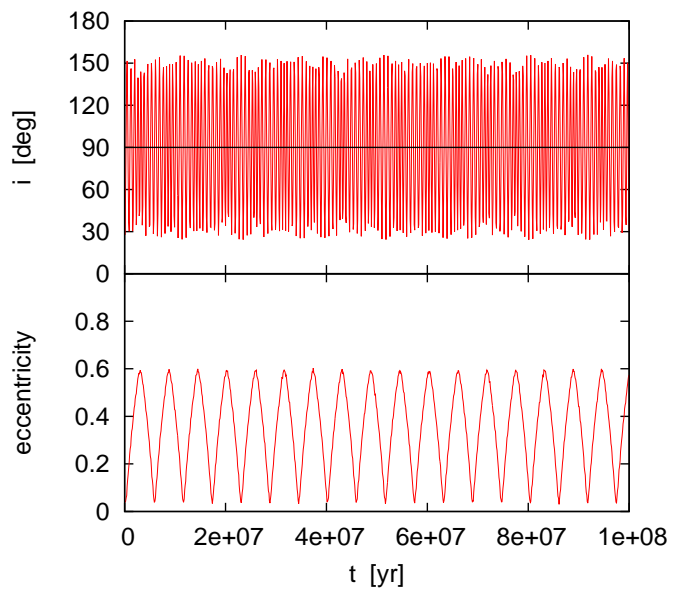}
\caption{
Evolution in time of the inclination $i$ (top panel) and the eccentricity $e$ (bottom panel) of a particular Type-F particle. The system
is composed of a Jupiter-mass planet with semimajor axis $a =$ 9.9 au, and eccentricity $e =$ 0.95 and a small body reservoir whose
members are all Type-F particles. Moreover, the particles undergo large amplitude of oscillation in eccentricity. The post IE orbital parameters
of this particle are $a_{1}$ $=$ 32 ua, $e_{1}$ $=$ 0.05, $i_{1}$ $=$ 33.1$\degr$, $\omega_{1}$ $=$ 170.9$\degr$, $\Omega_{1}$ $=$ 132$\degr$.
}
\label{fig:fig11}
\end{figure}

On the other hand, a general result obtained in our simulations indicates that the eccentricity $e$ of a Type-F particle evolves with a constant value or with librations of low-to-moderate amplitude (Fig.~\ref{fig:fig10}). In fact, such as \citet{Naoz2017} observed, the canonical specific momentum $J_{2} = \sqrt{1 - e^{2}}$ is conserved at quadrupole level of secular approximation, and thus the eccentricity of the test particle $e$ remains constant. However, those authors also showed that introducing the octupole level of approximation allows for variations of the test particle's eccentricity. It is worth noting that, in a few particular cases, our simulations produce Type-F particles whose eccentricity $e$ evolves undergoing large libration amplitudes of up to 0.6. In fact, this behavior can be seen in Fig.~\ref{fig:fig11}, which shows the evolution of the inclination $i$ and the eccentricity $e$ of a Type-F particle belonging to a system whose surviving giant planet has a semimajor axis $a=$ 9.9 au and an eccentricity $e =$ 0.95. A relevant feature of this system is that the surviving small body reservoir is composed of only Type-F particles whose eccentricities show an evolution similar to that observed in Fig.~\ref{fig:fig11}. This particular evolution of the eccentricity was also found in three Type-F particles associated to system whose surviving gas giant has a semimajor axis $a =$ 3.57 au and an eccentricity $e =$ 0.79.

\subsection{Orbital parameters of Type-F particles immediately after the instability event}

\begin{figure*}
\centering
\includegraphics[angle=270, width=0.98 \textwidth]{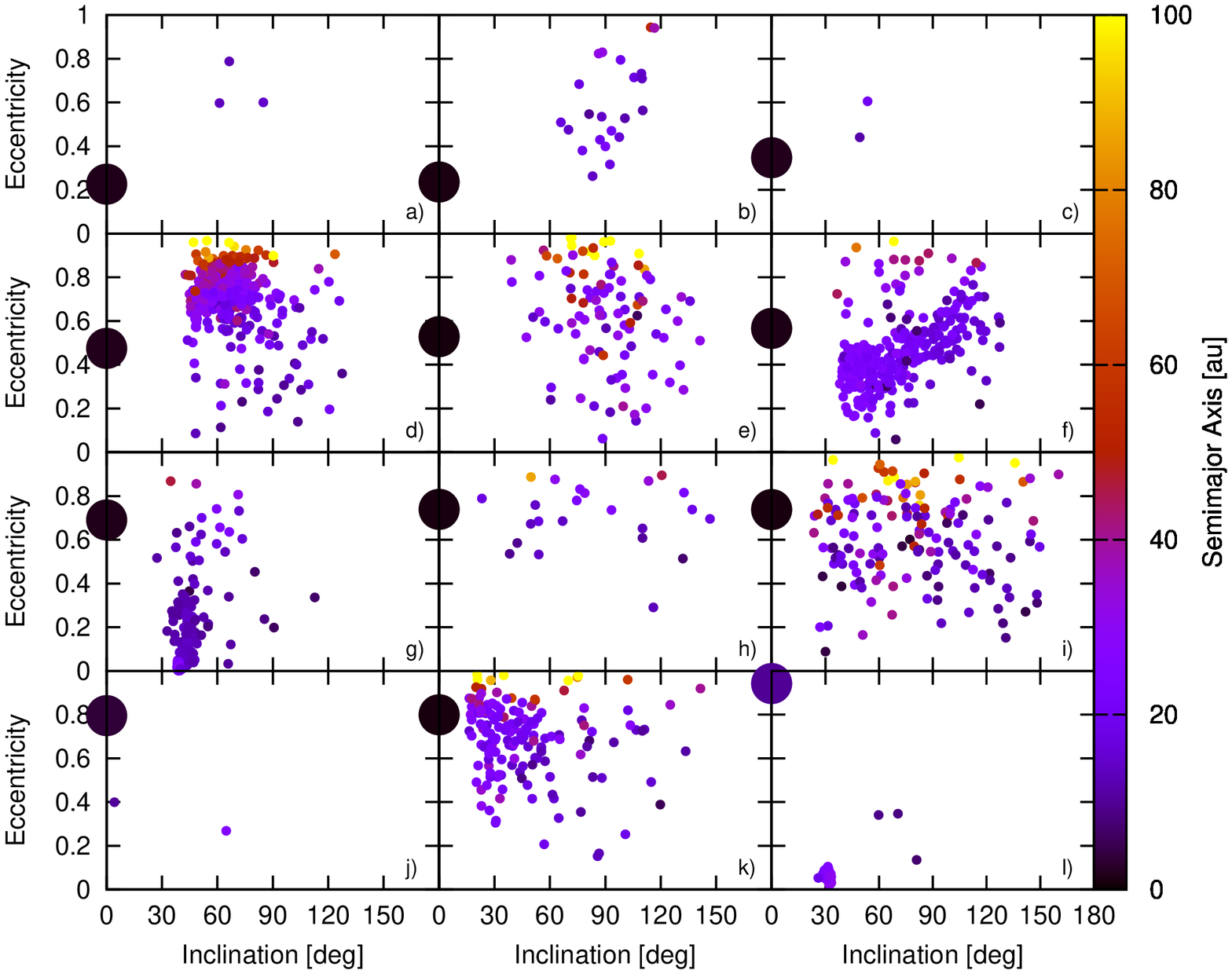}
\caption{
Post IE orbital parameters $i$ vs. $e$ of all Type-F particles and the eccentric inner perturber of the 12 simulations.
The scale color represents the post IE semimajor axis of the Type-F particles and of the surviving giant planet.
}
\label{fig:fig12}
\end{figure*}

We analyzed the orbital parameters immediately after the instability event (post IE orbital parameters) of Type-F particles of all our simulations with the goal to determine the parameter space that lead to the flipping of an orbit. In general terms, the post IE conditions are different from one to other simulation due to the strong planetary scattering event. Figure~\ref{fig:fig12} shows the post IE orbital parameters in a plane $e$ vs. $i$ for each of the 12 TFSs. Moreover, we apply as weighting factor to the post IE semimajor axis for each case by a color palette. We remark that the different panels are sorted by eccentricity of the surviving giant planet.

As shown in Fig.~\ref{fig:fig12}, we observed a wide diversity of post IE semimajor axes, eccentricities, and inclinations that lead to the flipping of the test particle's orbit. On the one hand, the test particle's orbit can flip for any value of the post IE eccentricity. In fact, we found Type-F particles with quasi-circular orbits (panels d,e,f,g,i,l), with moderated eccentricities (all panels), as well as with very high eccentricities close to unity (panels b,d,e,f,i,k) immediately after the instability event. On the other hand, it is worth noting that our simulations produced Type-F particles from prograde and retrograde orbits. However, it is important to remark that we found only two Type-F particles with post IE inclinations $i <$ 17$\degr$ in all our numerical simulations.

\begin{figure}[ht]
\centering
\includegraphics[angle=270, width=0.49\textwidth]{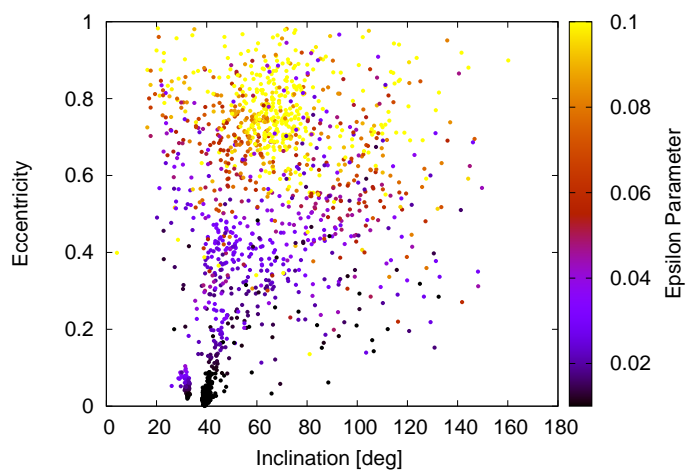}
\caption{
Post IE orbital parameters of all Type-F particles in a same plot. The color palette represents the post IE $\epsilon$ parameter.
}
\label{fig:fig13}
\end{figure}

In Figure~\ref{fig:fig13} we show the orbital parameter distribution immediately after the instability event of all Type-F particles observed in Fig.\ref{fig:fig12} in a plane $e$ vs. $i$. In this case, the color palette represents the $\epsilon$ parameter defined in Naoz et al. (2017), which is given by 
\begin{equation}
\epsilon = \frac{a_{\text{pla}}}{a_{\text{part}}}\frac{e_{\text{part}}}{(1-e_{\text{part}}^{2})}, 
\end{equation}
where $a_{\text{pla}}$ represents the semimajor axis of the giant planet, and $a_{\text{part}}$ and $e_{\text{part}}$ are the semimajor axis and eccentricity of test particle, respectively. To define such a color palette, the $\epsilon$ parameter was evaluated immediately after the instability event. As seen in Fig.~\ref{fig:fig13}, our results seem to suggest that the larger the value of post IE $\epsilon$, the larger the post IE eccentricity of the Type-F particle.

\subsection{Dependency of the particle's minimum inclination on the planet's eccentricity}
A detailed analysis of Fig. \ref{fig:fig12} allows us to suggest a strong correlation between the minimum inclination of a Type-F particle and the planet's eccentricity $e_{\text{pla}}$. From this, we decided to compute the minimum value of the inclination $i_{\text{min}}$ reached by the Type-F particles throughout the evolution for each of 12 panels of Fig. \ref{fig:fig12}. A notable feature derived from this analysis suggests that the minimum value of the inclination $i_{\text{min}}$ of the Type-F particles (which corresponds with $\Omega$ $=$ 90$\degr$ or 270$\degr$) decreases with an increase in the eccentricity $e_{\text{pla}}$ of the giant planet. This result can be observed in Fig.~\ref{fig:fig14}, in which each circle represents the values of $i_{\text{min}}$ and $e_{\text{pla}}$ associated to every panel of Fig.\ref{fig:fig12}. We added the analytical relation between the largest $i_{\text{min}}$ allowed for a test particle in the libration mode and the eccentricity of the giant planet, which was derived by \citet{Naoz2017} at the quadrupole level of secular approximation. Such a relation is illustrated as a green curve in Fig.~\ref{fig:fig14}. The post IE semimajor axis ratio $a_{\text{pla}}$/$a_{\text{part}}$ (top panel) and the post IE $\epsilon$ parameter (bottom panel) were aggregated as a weighting factor by a color palette. When the semimajor axis ratio is high, higher order effects in the secular approximation may play an important role. Moreover, if the value of $\epsilon$ parameter is high, the planet and the particle can approach and non-secular effects may play a relevant role in the dynamical evolution of the particle. As the reader can see, all circles located below the green analytical curve in Fig.~\ref{fig:fig14} have high values of $\epsilon$ parameter (i.e., $\epsilon \gtrsim$ 0.1), so that non-secular effects may be important in the dynamical behavior of the particles. Moreover, some of circles located below the analytical curve show high values of the semimajor axis ratio $a_{1}$/$a_{2}$ (i.e., $a_{1}$/$a_{2} \gtrsim$ 0.1), for which higher order effects in the secular approximation may also be relevant in the dynamical evolution. However, it is worth noting that such points do not show relevant deviations from the analytical predictions.

\section{Discussion and conclusions}

Observational and theoretical considerations have suggested that a large fraction of the known extrasolar planetary systems have
undergone a period of large-scale dynamical instability involving planet-planet scattering events.

Planetary scattering events involving gaseous giants naturally lead to violent scenarios of dynamical evolution. In fact, in such
events, one o more giant planets are ejected from the system on hyperbolic trajectories, while the surviving planets (if they would
exist) can achieve highly eccentric and inclined orbits.

A very important topic concerning these systems is the dynamical evolution of the remnant small body reservoirs. In fact,
depending on the time duration of the planet-planet scattering event and the orbital parameters of the surviving planet (or planets),
the remnant populations of small bodies will show different dynamical features.

In the present work, we analyzed the dynamical evolution of planetary systems that harbor three Jupiter-mass planets close to their stability limit around stars of 0.5 M$_{\odot}$. In particular, we focused on analyzing the planetary configurations and the dynamical
features of the icy body reservoirs produced from planetary scattering events in systems with a single surviving Jupiter-mass planet. Due
to the chaoticity of the dynamics involved in such events, we carried out a total of 21 N-body simulations making use of RA15 version of
the RADAU numerical integrator \citet{Everhart1985}, which is included in the Mercury code \citet{Chambers1999}. Such an integrator is a very
useful numerical tool due to its stability and precision when dealing with gravitational encounters with Jupiter-mass bodies.

Our N-body simulations have lead to a wide diversity of planetary architectures and small body reservoirs. On the one hand, the Jupiter-mass planet that survived to the planetary scattering event adopted values for the semimajor axis and eccentricity of 0.5 au to 10 au, and 0.01 to 0.94, respectively. On the other hand, the small body populations showed very different dynamical features in the evolution of the test particles. In fact, our simulations produced particles with prograde (Type-P) and retrograde (Type-R) orbits, as well as ``flipping particles'' (Type-F), whose orbits flip from prograde to retrograde or vice versa along their evolution.

The study of outer minor bodies whose orbits flip due to the action of an inner eccentric giant planet results to be very important since it could represent a very common scenario between the extrasolar planetary systems in the Universe. In the present research, we found very interesting dynamical features in the evolution of the Type-F particles. We also derived several properties concerning the inclination $i$ and the ascending node longitude $\Omega$ of the Type-F particles, which are the following:

\begin{itemize}
\item[1-] Oscillations of $\Omega$ around 90$^{\circ}$ or 270$^{\circ}$: this represents a necessary and sufficient condition for the flipping of an orbit,
\item[2-] Equality between the libration periods of $i$ and $\Omega$,
\item[3-] Out-of-phase periodicity of $i$ and $\Omega$ by a quarter of period,
\item[4-] The larger the libration amplitude of $i$, the larger the libration amplitude of $\Omega$.
\end{itemize}
We also inferred several properties concerning the eccentricity of the flipping particles. In fact, our simulations showed that the flipping of an orbit can be produced for values from very low and very high of the particle's eccentricity, which, in general terms, evolves with a constant value or with librations of low-to-moderate amplitude. However, we observed a few Type-F particles with large libration amplitudes of up to 0.6 associated to the eccentricity.

We also carried out an analysis concerning the orbital parameters of the Type-F particles immediately after the dynamical instability event (post IE parameters). Our results suggest that the orbit of a test particle can flip for any value of post IE eccentricity. However, we found only two Type-F particles with post IE inclinations $i <$ 17$\degr$ in all our numerical simulations.

Finally, our study indicates that the minimum value of the inclination $i_{\text{min}}$ of the Type-F particles reached throughout the evolution in a given system decreases with an increase in the eccentricity $e_{\text{pla}}$ of the giant planet. In fact, the larger the eccentricity of the giant planet, the lower the minimum value of the inclination of the Type-F particles in a given system.

\begin{figure}[ht]
\centering
\includegraphics[angle=270, width=0.48\textwidth]{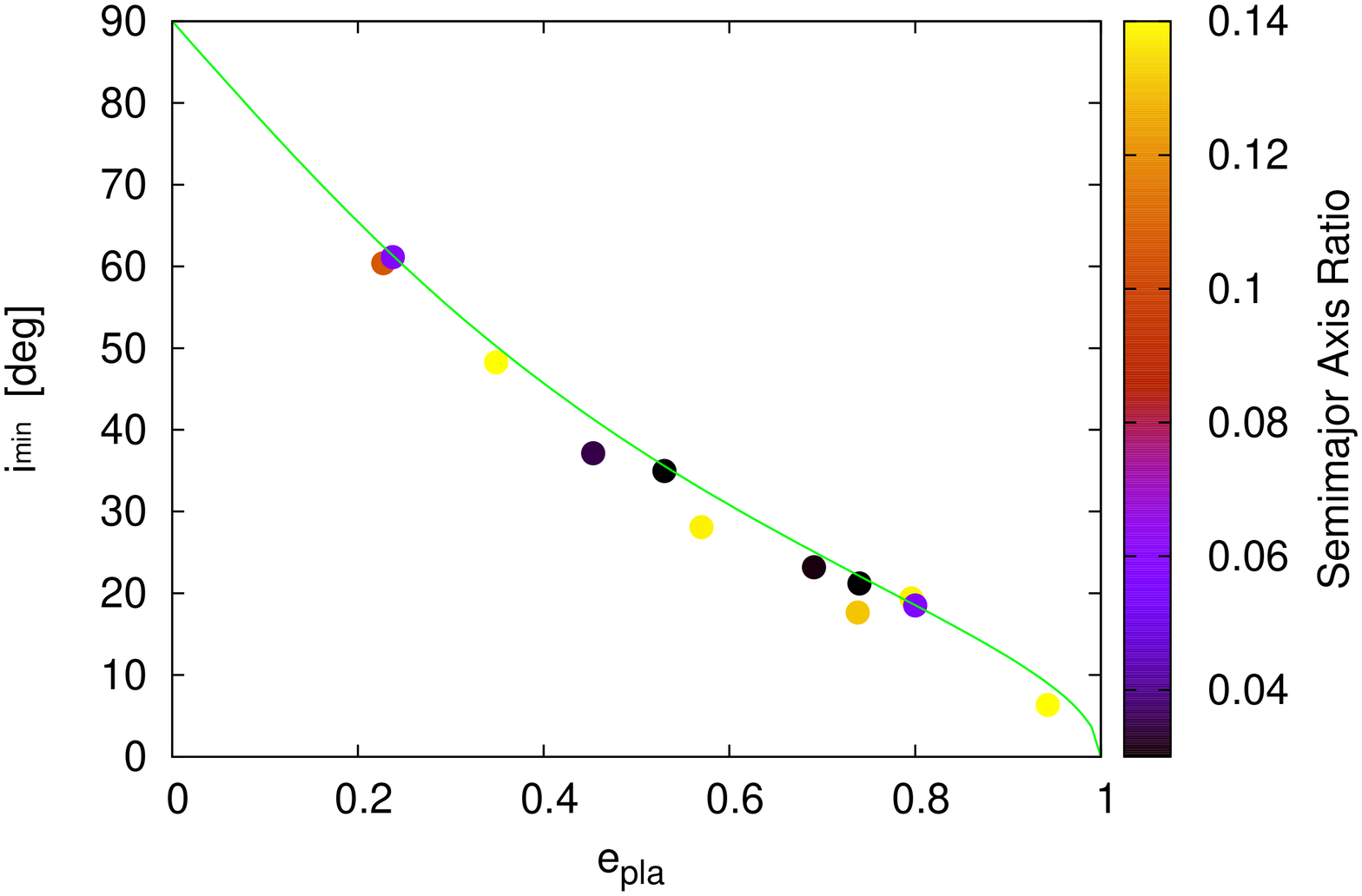}\\
\includegraphics[angle=270, width=0.48\textwidth]{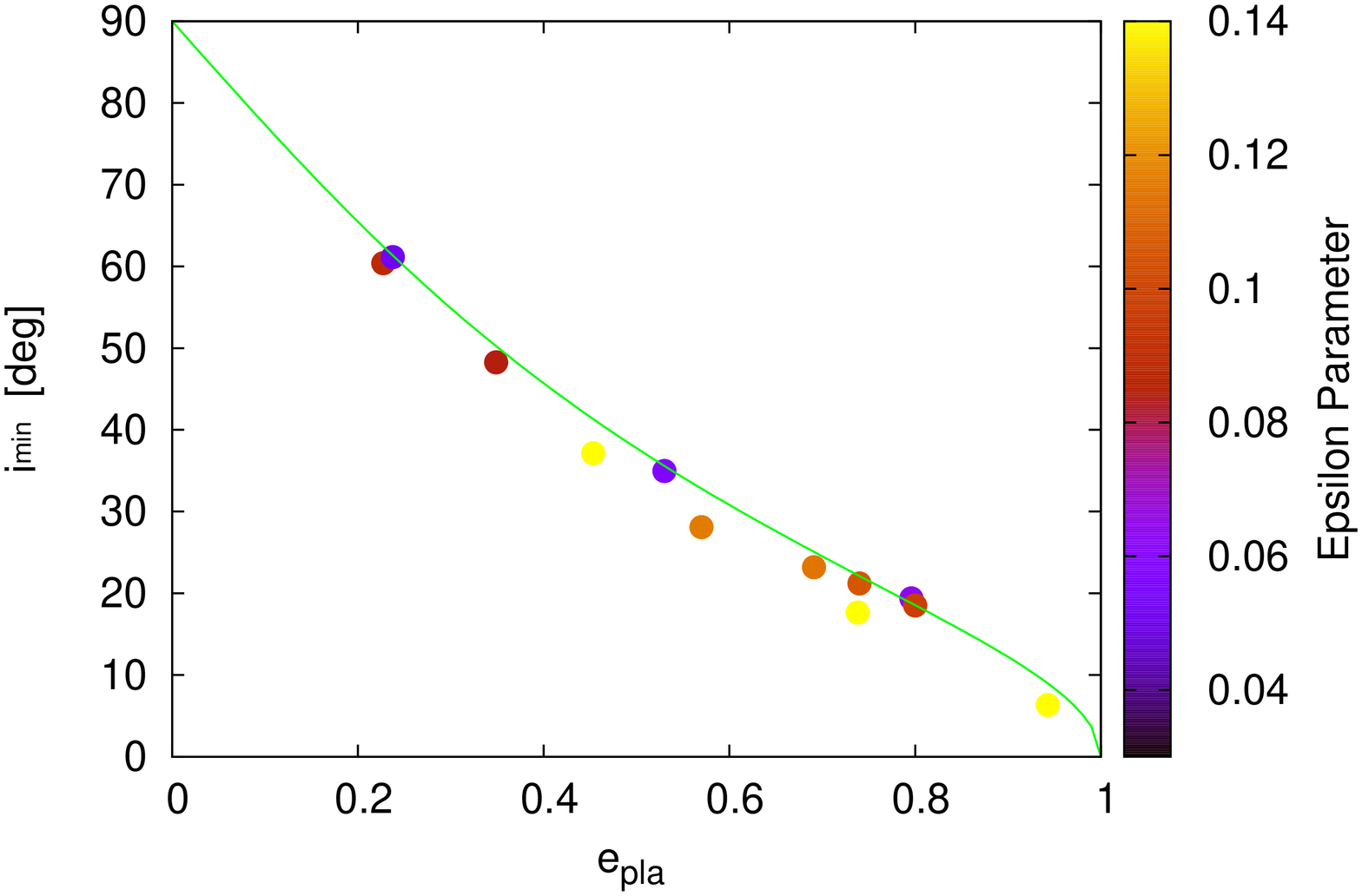}
\caption{
Minimum inclination of Type-F particles vs. eccentricity of the surviving planet. A weighting factor given by the post IE semimajor axis ratio (top panel) and the post IE $\epsilon$ parameter (bottom panel) is included. Our results indicate that the larger eccentricity of the surviving planet, the lower the minimum inclination of Type-F particles. The green curve represents the analytical relation obtained by \citet{Naoz2017}.
}
\label{fig:fig14}
\end{figure}

Our research provides evidence for a correlation between the orbital properties of an outer flipping particle and the eccentricity of an inner massive planet. Our simulations show that only it was possible to find Type-F particles in those systems with a Jupiter-mass planet with an eccentricity $e >$ 0.2. However, it is worth noting that the presence of an eccentric giant planet does not guarantee the existence of a Type-F particle in the system (see Fig.~\ref{fig:fig3}). From Fig.~\ref{fig:fig14}, it is interesting to see that the larger the eccentricity of the inner giant planet, the wider the range of inclinations of an outer flipping test particle. Thus, an inner Jupiter-mass planet with an eccentricity $e <$ 0.2 is also able to produce outer flipping particles but the range of possible inclinations is smaller than that associated to more eccentric planets.

Our numerical results concerning the dynamical evolution of Type-F particles are in agreement with the predictions derived by Naoz et al. (2017), who studied the secular evolution of an outer test particle under the effects of an inner eccentric planet up to the octupole level of approximation. 
It is worth noting that the slight deviations observed between our numerical results and \citet{Naoz2017}'s predictions are due to the influence of non-secular and higher order secular effects. This allows us to understand the large oscillations associated to the particle's eccentricity in Fig.~\ref{fig:fig11}, as well the slight deviations of the circles from the green analytical curve in Fig.~\ref{fig:fig14}, which was derived by \citet{Naoz2017} up to quadrupole level of approximation.

We note that the general relativity could play an important role in the dynamical behavior of these systems. Recently, \citet{Naoz2017} suggested that the general relativity effect could suppress the inclination excitation of the outer test particle, depending on its dynamical properties as well as on the orbital parameters of the inner perturber. These effects will be include in a forthcoming paper.

The research presented here describes the effects generated by an inner eccentric Jupiter-mass planet on an outer small body population resulting from a scattering event involving three Jupiter-mass planets. Previous studies such as \citet{Raymond2008}, \citet{Raymond2013}, and \citet{Marzari2014b} show that the scattering event is sensitive to the initial mass distribution of the planets. From such results, we consider that the analysis developed in the present work could to be extended to other different initial configurations for the planets. In fact, it would be interesting to investigate systems composed of disturbers with different physical and orbital properties to strengthen our understanding about the dynamical evolution of outer small body populations.

It is important to remark that the production of dust is a topic of special interest in the present study. In fact, the dust produced by the reservoirs formed in our simulations may be detectable at IR wavelengths. In fact, a study about the dust production requires an analysis concerning the collisional evolution of our reservoirs. The collisional evolution models are very sensitive to several properties of the small body population, such as the impact velocity, the intrinsic collision probability, the composition of the bodies, slopes and breaks associated to the size distribution, size of the largest body of the population, between others \citep{Petit1993,OBrien2005,Bottke2005,deElia2007a,deElia2007b,deElia2008,deElia2010}. A comparison between our numerical results and observations concerning the emission of could dust in systems with an eccentric giant planet should help us to improve our theoretical models of formation and evolution of planetary systems. We are now developing a detailed study about the production of dust in our systems, which will be discussed in a forthcoming paper.

A detailed analysis of the influence of an inner eccentric Jupiter-mass perturber on an outer small body reservoir represents a very important work for the understanding of several very peculiar dynamical features that could be evident in many known extrasolar planetary systems.

\begin{acknowledgements}
This work was partially financed by CONICET and Agencia de Promoci\'on Cient\'{\i}fica, through the PIP 0436/13 and PICT 2014-1292. We acknowledge the financial support by FCAGLP for extensive use of its computing facilities and Dr. Tabar\'e Gallardo for valuable discussions and useful comments about the numerical runs. We also thank anonymous referee for his/her helpful and constructive reviews. SN acknowledges partial support from a Sloan Foundation Fellowship. GL is supported in part by the Harvard William F. Milton Award.
\end{acknowledgements}

\bibliographystyle{aa}          
\bibliography{zanardi} 

\end{document}